\documentclass{elsart}
\usepackage{epsfig}
\begin{document}
\title{Information theory in high energy physics \\
(extensive and nonextensive approach)} 
\author{F.S.Navarra$^{a}$, O.V.Utyuzh$^{b}$, G. Wilk$^{b}$, 
Z.W\l odarczyk$^{c}$} 
\address{$^{a}${\it Instituto de F\'{\i}sica, Universidade de S\~{a}o
               Paulo\\
               CP 66318, 05389-970 S\~{a}o Paulo, SP, Brazil; e-mail:
               navarra@if.usp.br}\\
         $^{b}$The Andrzej So\l tan Institute for Nuclear Studies\\
                Ho\.za 69; 00-689 Warsaw, Poland; e-mail: wilk@fuw.edu.pl\\
         $^{c}$Institute of Physics, \'Swi\c{e}tokrzyska Academy,\\
               \'Swi\c{e}tokrzyska 15; 25-406 Kielce, Poland; e-mail:
               wlod@pu.kielce.pl\\ 
 \today}
{\scriptsize Abstract: The application of information theory approach
(both in its extensive and nonextensive versions) to high energy 
multiparticle production processes is discussed and confronted with
experimental data on $e^+e^-$ annihilation processes, $pp$ and
$\bar{p}p$ scatterings and heavy ion collisions. 


}
\section{Introduction}

High energy multiparticle production processes are most naturally
described by statistical models (see \cite{STAT,HAG} for a historical
background and \cite{NEWSTAT,NEWSTAT1} for the most recent developments).
Their results are usually interpreted in the thermodynamical 
sense, with temperature $T$ and chemical potential $\mu$ entering 
with their usual meaning. However, it has been recognized that in 
this branch of physics one very frequently encounters "thermal-like" 
form of distributions in some variable $x$, say $\propto \exp(-x/T)$, 
without system under consideration being in any kind of thermal
equilibrium \cite{RES}. It is enough that out of the huge number 
of produced secondaries only some part is registered by detectors,
and out of them only one or two are selected for final scrutiny.
The averaging emerging this way is then equivalent to the action of
some "heat bath" characterized by parameter $T$. Once this is
realized it is then obvious that there are situations in which such
"heat bath" is more complicated (for example nonextensive) and needs 
additional parameter(s) to be described properly \cite{Q,Q1,Q2}. In this way
the nonextensive Tsallis statistics characterised by the non-extensivity 
parameter $q$ enters in a natural way \cite{T,WW,WWq,BeckC}. On the
other hand, such apparently "thermal-like" behaviour of some distributions
arises very naturally in many physical applications of information
theory (MaxEnt) \cite{INFO} (with Shannon form of the corresponding
information entropy, its nonextensive form uses Tsallis entropy
characterised by the same parameter $q$ as mentioned above, such
that for $q\rightarrow 1$ one recovers the usual Shannon entropy). 

\section{Results}

The usefulness of information theory is most obvious in situations 
when one has to "guess" the most probable (and least biased) distribution 
$p(x)$ of some quantity $x$ using only limited amount of information 
given in terms of finite number $n$ of some observables:
$R_{k=1,\dots ,n} = \langle R_k(x)\rangle$ ($p(x)$ is normalized to
unity and average $\langle \dots \rangle$ is with respect to $p(x)$ 
in extensive and to $[p(x)]^q$ in nonextensive approaches,
respectively \cite{INFO,T}. In high energy multiparticle production
processes it allows to find, in a maximally model independent way,
{\it the real  information content of the experimental data
considered} \cite{Chao,MaxEnt,OMT,NC,qMaxEnt}. From \cite{Chao} we
know therefore that: $(i)$ -  particles produced in high energy
collisions are mostly located in one dimensional phase space, i.e.,
they have limited transverse (with respect to collision axis) momenta
$p_T$ ($\langle p_T\rangle$ is finite) and are fully characterized by
their longitudinal momenta $p_L=\mu_T \sinh y$ \footnote{Here
$\mu_T=\sqrt{\mu^2 + \langle p_T\rangle^2}$ is the so called
transverse mass of particle with mass $\mu$ and transverse momentum
$p_T$ and $y$ denotes rapidity of the particle, variable defined in
such way that its energy is $E=\mu_T \cosh y$.}; $(ii)$ -  only
fraction $K\in (0,1)$ (called {\it inelasticity}) of the initially
available energy $W$ is converted into produced particles. As was
shown recently by us \cite{qMaxEnt} the fact that multiplicity
distribution of produced secondaries is no longer of Poissonian type
but follows much broader Negative Binomial (NB) form can be accounted
for {\it only} by using nonextensive approach with $q$ given by the
parameter $k$ of NB, $q=1+1/k$, which represents the amount of
dynamical fluctuations in the number of produced secondaries
\cite{qMaxEnt}. 

We are usually interested in single particle rapidity
distributions, which in the information theory approach are given by
\begin{equation}
p(y)\, =\, \frac{1}{N}\frac{dN}{dy}\, =\, \frac{1}{Z_q} \exp_q 
       \left( - \beta_q \cdot \mu_T \cosh y \right). \label{eq:py}
\end{equation}
This form is identical with that used in statistical models but now
$Z_q$ and $\beta_q$ are no longer {\it free parameters} to be fitted
when comparing with experimental data but instead are given by the
normalization condition and energy conservation constraint,
\begin{equation}
\int_{-Y_m}^{Y_m}\, dy\, p(y)\, =\, 1 \qquad {\rm and}\qquad 
\int_{-Y_m}^{Y_m} \, dy\, \mu_T\cdot \cosh y \cdot [p(y)]^q\, =\,
\frac{\kappa_q \cdot W}{N} \label{eq:energy}
\end{equation}
(where $\pm Y_m$ are maximal rapidities available in rest frame of
hadronizing source, see \cite{MaxEnt,qMaxEnt} for details).
There are therefore two parameters: nonextensivity $q$ responsible 
for dynamical fluctuations and $q$-inelasticity $\kappa_q$ to be deduced
directly from data essentially in a {\it model independent 
way} \footnote{~~Here $\exp_q\left( x/\Lambda\right) = \left[1 +
(1-q)x/\Lambda\right]^{1/(1-q)} \stackrel{q \rightarrow
1}{\Rightarrow} \exp (x/\Lambda)$). Possible question concerning
physical meaning of $\kappa_q$ for $q\neq 1$ case is solved by
noticing \cite{qMaxEnt} that the inelasticity parameter in this case,
which has physical meaning, is $K_q = \kappa_q/(3-2q)$.}. 
\begin{figure}[ht]
  \begin{minipage}[ht]{57mm}
    \centerline{
        \epsfig{file=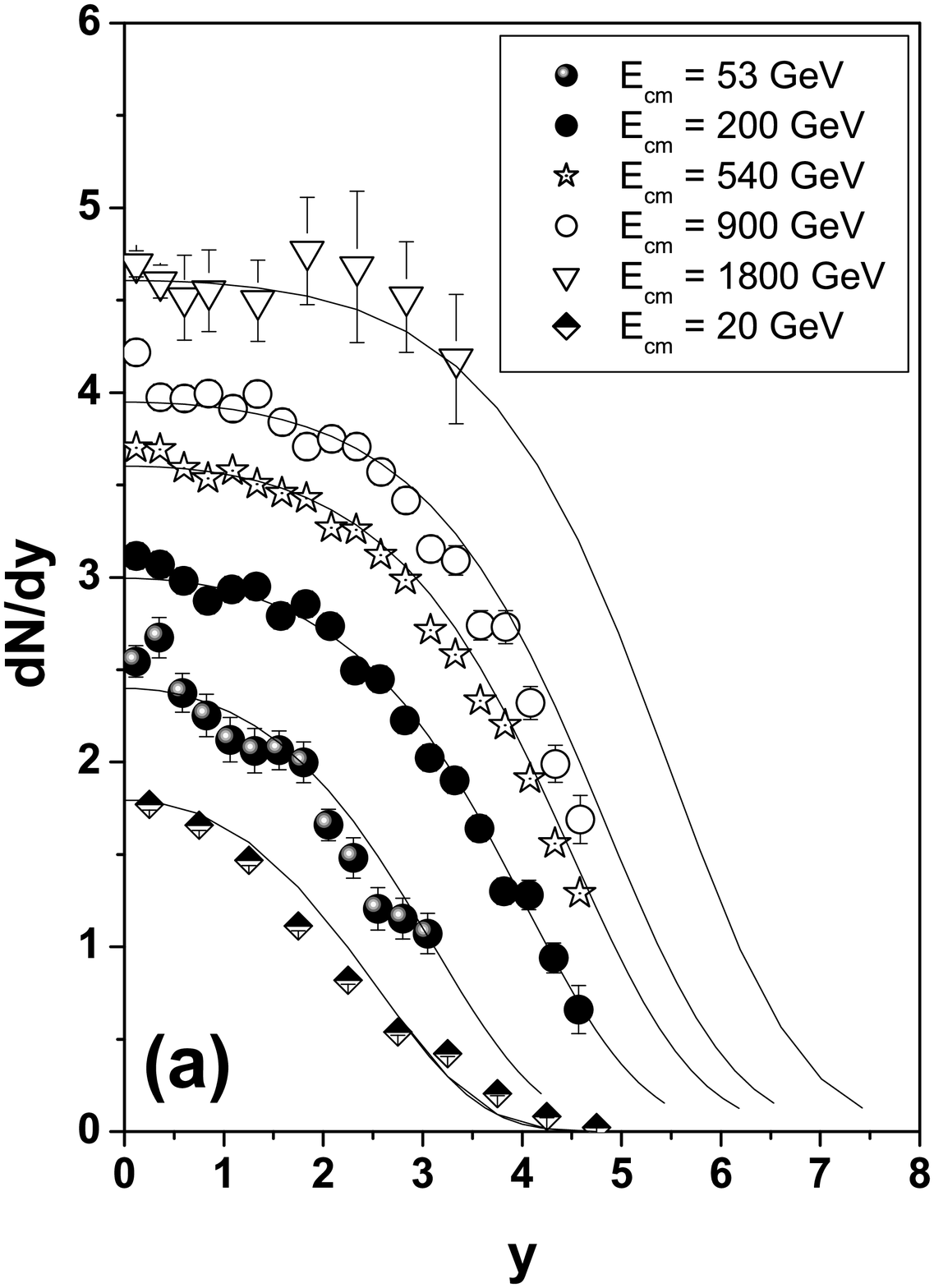, width=55mm}                             
     }
  \end{minipage}
\hfill
  \begin{minipage}[ht]{57mm}
    \centerline{
        \epsfig{file=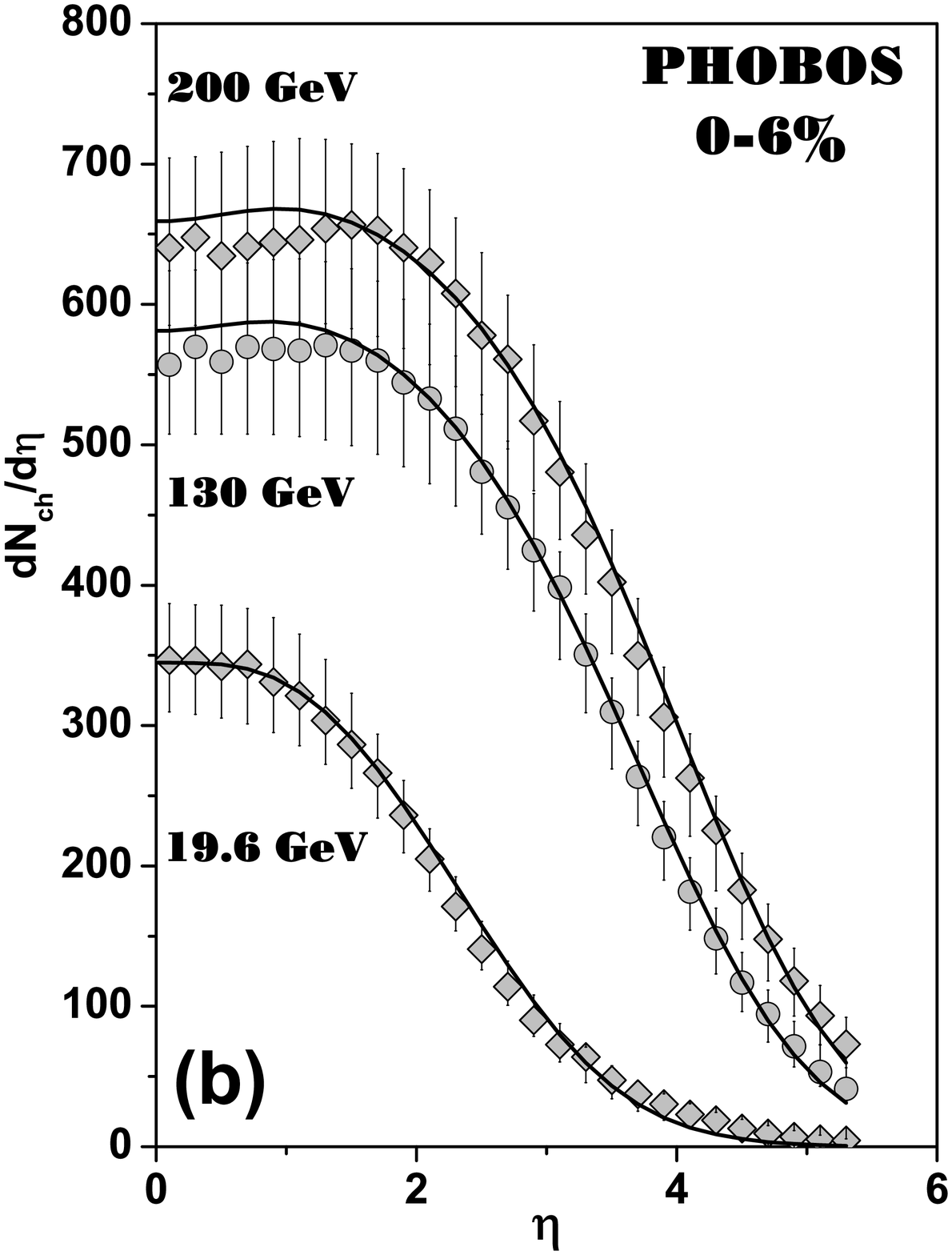, width=55mm}
     }
  \end{minipage}
\hfill
  \begin{minipage}[ht]{57mm}
    \centerline{
       \epsfig{file=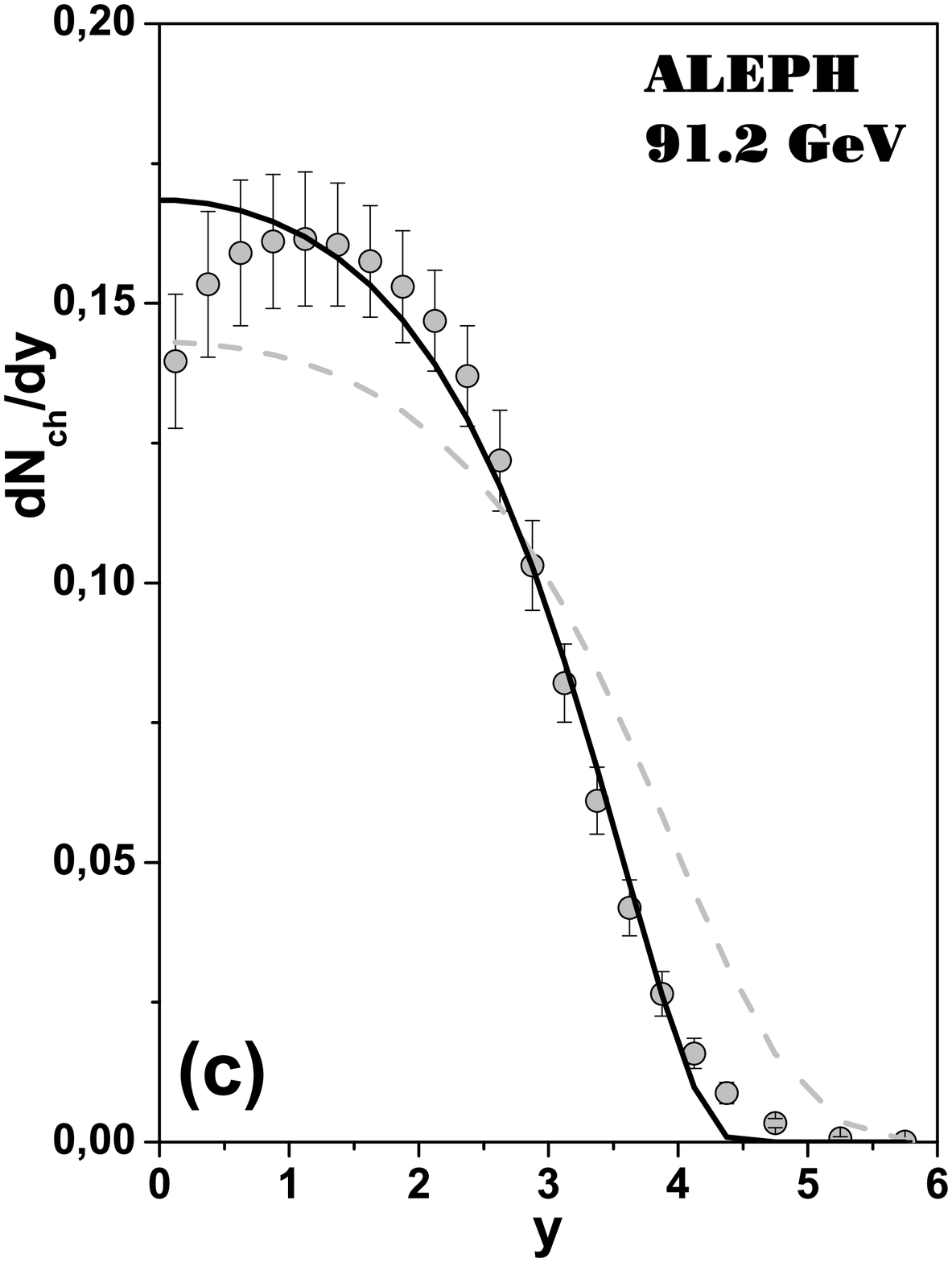, width=55mm}
     }
  \end{minipage}
  \caption{
\footnotesize Examples of applying eq. (\ref{eq:py}) to: $(a)$
rapidity spectra for charged pions produced in $pp$ and $\bar{p}p$
collisions at different energies \cite{Data1,Data1a,Data2,Data3};
$(b)$ similar data obtained for the most central $Au+Au$ collisions
\cite{Data4}; $(c)$ rapidity spectra measured in $e^+e^-$
annihilations at $91.2$ GeV \cite{Data5} (dotted line is for $K_q=1$
and $q=1$ whereas full line is our fit with $K_q=1$ and $q=0.6$).}
  \label{fig:Fig1}
\end{figure}
This method works perfectly well for $pp$ and $\bar{p}p$ collisions
\cite{qMaxEnt}, cf. Fig. \ref{fig:Fig1}a (where it provides us with
the first model independent estimations of energy dependence of the
mean inelasticity parameter and with its distribution). Here we show
that it works also quite well for similar data on rapidity
distributions obtained in $Au+Au$ collisions, cf. Fig.
\ref{fig:Fig1}b. The $Au+Au$ data are for the most central events
(covering collisions proceeding with impact parameter range $0-6\%$).
They can be fitted choosing $K_q=1$ and then $q= 1.29$, $1.26$ and
$1.27$ for energies $19.6$, $130$ and $200$ GeV, respectively
(the $q$-inelasticity was  therefore equal to $\kappa_q = 0.42$,
$0.48$ and $0.46$)\footnote{~~~Actually data in Fig. \ref{fig:Fig1}b.
are presented for the so called pseudorapidity $\eta$ defind not by
energy $E$ and longitudinal momentum $p_L$ as is the case of
rapidity $y$ but by total momentum $p$ and longitudinal momentum
$p_L$ instead.}. The most interesting are, however, results for
$e^+e^-$ annihilations (cf. Fig. \ref{fig:Fig1}c) for which, by
definition, $K_q = 1$ (i.e., always all energy of initial leptons is
available for the production of secondaries) and which can  be fitted
{\it only} with $q<1$ (in our case $q=0.6$).

This point deserves closer scrutiny. One could argue that because fit
in Fig. \ref{fig:Fig1}c is not perfect (there are some discrepances
for small rapidities and there is a tail at large values of $y$)
there is nothing to be said before they are not addressed. But
results for $q=1$ clearly show that these discrepancies are not
connected with the particular value of $q$ but rather with some
additional mechanisms operating here action of which would, however,
change our results only slightly (for example, a possibility of two
rather than one source or $y$-dependent $\langle p_T\rangle$, as
mentioned already in \cite{Next}). With the above reservations let us
then take a closer look at the possible origin of $q<1$. We have
already encountered similar situation when in \cite{NC} we have
fitted single particle distributions assuming implicitely that
$K_q=1$ and discovering then that one could get fairly good agreement
with data for $q<1$ only. That was because in this case only $q<1$
leads effectively to cutting-off a part of the phase space (once it
is taken too big) mimicking therefore action of the inelasticity
parameter (cf. \cite{qMaxEnt}). On the other hand, we know that $q\neq
1$ signals presence of fluctuations in the system \cite{WW,WWq,BeckC}
and is given by normalized variations of these fluctuations, in our
case they would be fluctuations of temperature $T=1/\beta$ parameter.
So far it was widely discussed only for the $q>1$ case
\cite{WWq,BeckC} but  formally it covers the $q<1$ case as well, cf.
\cite{WWq}. However, in this case temperature $T$ does not reach an
equilibrium state because in this case
\begin{equation}
T\, =\, T_0\, -\, (1-q) E \label{eq:Tq}
\end{equation}
instead reamining constant, $T=T_0$, as is the case for $q>1$. In
this case we have a kind of dissipative transfer of energy from the
region where (due to fluctuations) the temperature $T$ is higher (for
example, in our case from the quark ($q$) and antiquark ($\bar{q}$)
jets formed  in the first $e^+e^- \rightarrow q + \bar{q}$ to gluons
and $q\bar{q}$ pairs and later on to finally observed hadrons). It
means therefore that $q<1$ signals that in the reaction considered,
where $K_q=1$ and we have to account for the whole energy exactly,
conservation laws start to be important and it is not possible for
stationary state with constant final temperature to develop but
temperature $T$ depends on the energy\footnote{~~~~Actually, when
fluctuations depend on it in the same way the relative variance
$\omega$ remains constant and $q=1-\omega$, cf. \cite{WWq}.} and for
large energies tends to zero (notice that from eq. (\ref{eq:Tq}) one
has limitation on the allowed energy of the secondaries: $E \le
T_0/(1-q)$)\footnote{~~~~~~~~Notice that analysis of $p_T$ distribution
in the same process using a kind of $q$-version of Hagedorn model
\cite{Bed} reports $q>1$ instead. This is, however, what we call
$q_T$ in \cite{Next} and this is different from $q$ considered here
(called $q_L$ in \cite{Next}, where we compared both $q$'s for
$p\bar{p}$ collisions). The reason for such different behaviour is
that $p_T$'s considered there are essentially not influenced by
conservation laws but, in our language here,  reflect instead a kind
of stationary state with $q>1$ and energy independent $T$.}.

\section{Summary}

We have demonstrated that information theory (know also as MaxEnt
method) can play very important role in high energy physics providing
simple, highly model independent, estimations of single particle
distributions and allowing to reliably estimate the amount of
information provided by experimental data. Its nonextensive version
extends applicablity range of MaxEnt by including also some intrinsic
fluctuations in the hadronizing system visible as broadening of the
multiplicity distributions.  As was shown here, single particle
distribution data on all kinds of collisions, starting form very  
elementary $e^+e^-$ annihilations, via $pp$ and $\bar{p}p$ collisions,
and ending with the most complicated heavy ion scatterings, can be
described by formula (\ref{eq:py}) depending on very limited number
of parameters: inelasticity $K$ and nonextensivity parameter $q$.
Fairly good fits were obtained in all cases. To proceed further one
should concentrate now on these parts of the phase space where
discrepancies occur and then introduce {\it one-by-one} some
additional hypothesis and check whether they lead to the better
agreement with data (see, for, example \cite{Next}). Such approach
allows to avoid assumptions, which whereas looking promising, are not
justified.  

\noindent                            
{\bf Acknowledgments:} Partial support of the Polish State Committee
for Scientific Research (KBN) (grant 2P03B04123 (ZW) and grants
621/E-78/SPUB/CERN/P-03/DZ4/99 and 3P03B05724 (GW)) is acknowledged.


\end{document}